\documentclass[journal]{IEEEtran}

\usepackage{graphicx,subfigure,amsmath,amssymb,amsfonts,cite,float,color,longtable,rotating,diagbox,booktabs,array}

\ifCLASSINFOpdf
\else
\fi

\begin{document}

\title{Robust Secure Transmission of Using Main-Lobe-Integration Based Leakage Beaforming in Directional Modulation MU-MIMO Systems}

\author{Feng Shu, Wei Zhu,~Xiangwei~Zhou,~Jun Li,~and~Jinhui Lu
\thanks{This work was supported in part by the National Natural Science Foundation of China (Nos. 61472190, 6147210, 61501238, and 61271230), the Open Research Fund of National Key Laboratory of Electromagnetic Environment, China Research Institute of Radiowave Propagation (No. 201500013), and the open research fund of National Mobile Communications Research Laboratory, Southeast University, China (No. 2013D02).}
\thanks{Feng Shu, Wei Zhu, Jun Li, and Jinhui Lu are with the School of Electronic and Optical Engineering, Nanjing University of Science and Technology, 210094, CHINA. Email: \{shufeng, wei.zhu, jinhui.lu, Jun.li\}@njust.edu.cn}.
\thanks{Xiangwei Zhou is with the Division of ECE, Louisiana State University, 102 South Campus Drive, Baton Rouge, LA 70803, USA.}
\thanks{Feng Shu is also with the National Key Laboratory of Electromagnetic Environment, China Research Institute of Radiowave Propagation, Qingdao 266107, China, and National Mobile Communications Research Laboratory, 210096, Southeast University, China.}
}

\maketitle

\begin{abstract}
In the paper, we make an investigation of robust beamforming for secure directional modulation in the multi-user multiple-input and multiple output (MU-MIMO) systems in the presence of direction angle measurement errors. When statistical knowledge of direction angle measurement errors is unavailable, a novel robust beamforming scheme of combining main-lobe-integration (MLI) and leakage is proposed to simultaneously transmit multiple different independent parallel confidential message streams to the associated multiple distinct desired users. The proposed scheme includes two steps: designing the beamforming vectors of the useful confidential messages and constructing artificial noise (AN) projection matrix. Here, in its first step, the beamforming vectors for the useful confidential messages  of desired user $k$ are given by minimizing the useful signal power leakage from main-lobe of desired user $k$ to the sum of  main-lobes of the remaining desired directions plus main-lobes of all eavesdropper directions. In its second step, the AN projection matrix is constructed  by  simultaneously maximizing the  AN power leakage to all eavesdropper directions such that all eavesdroppers are disrupted seriously, where AN is viewed by the transmitter as a useful signal for eavedroppers. Due to independent beamforming vectors for different desired users,  a more secure transmission is achieved. Compared with conventional non-robust methods, the proposed method can provide a significant improvement in bit error rate along the desired directions and secrecy-sum-rate towards multiple desired users without needing statistical property or distribution of angle measurement errors.
\end{abstract}


\IEEEpeerreviewmaketitle

\section{Introduction}
The privacy and security of information transmission are extremely important for wireless communications and networking \cite{Hong2013, Zou2016, Huang2017, Liu2016, Ning2017_1, Ning2017_2}. Due to the broadcasting nature and lack of physical boundaries of wireless transmission, the information is readily intercepted by unauthorized users, and the growing cyber criminal events in mobile terminals have exposed the enormous hidden risks in wireless communications and networking. Traditional encryption techniques can only provide computational security and rely heavily on the complexity of their keys, that are very easy to be cracked if an efficient method to solve the corresponding mathematical problem is found \cite{Hong2013, Zou2016, Huang2017, Liu2016}. In the emerging social aware network, in order to take full advantage of wireless network resources, several incentive mechanisms based on credit or friendship had been presented to stimulate selfish users to forward data for other terminals, which will result in more serious security problems if the useful messages are intercepted by malicious users\cite{Ning2017_1, Ning2017_2}. In 1975, Wyner first proposes the remarkable wiretap channel model and lays the foundations of information theory of physical-layer security communications in \cite{Wyner1975, Leung1978, Csiszar1978}. In recent years, the physical-layer security has ever been becoming a promising research field in wireless communications and networking by exploiting the physical-layer characteristics of wireless channels \cite{Bloch2008, Mukherjee2011, Zou2015_2, Zou2015_1, Chen2016, Zhao2016_2}.

As a novel physical-layer security technology in wireless communications and networking, directional modulation (DM) has been attracting widespread attention and research activities from both academia and industry. In \cite{Babakhani2008}, the authors introduce a technique of near-field direct antenna modulation (NFDAM), changing the antenna boundary conditions at the symbol rate thereby modulating the phase and amplitude of the antenna pattern. A simplified structure of DM synthesis relying on actively excited elements in phased arrays is described in \cite{Daly2009, Daly2010}. Subsequently, the artificial-noise-aided security emerges and is applied to confidential transmission \cite{Goel2008, Zhou2009, Zhao2016_1}, which enables the transmitter to transmit artificial noise (AN), as an interference signal, and confidential messages together, such that the AN is only used to interfere with the eavesdroppers without affecting the legitimate receivers. The algorithm in \cite{Mukherjee2011} allocates that all the remaining available transmit power to transmit AN in the case that the desired receivers are guaranteed to achieve a target signal-to-interference-plus-noise-ratio (SINR). Based on the previous work, the authors in \cite{Ding2014, Ding2015_1} initiate the DM research on the baseband signal processing and propose to add the orthogonal vector, which can be chosen and updated in the null space of channel vector at the desired direction, to the transmitted baseband signal as AN \cite{Ding2014, Ding2015_1}. An orthogonal-vector-approach-based synthesis of multi-beam DM follows immediately from \cite{Ding2015_2}. The methods in \cite{Ding2014} and \cite{Ding2015_1} are shown to perform very well for the perfect direction angles but be quite sensitive to the measurement errors of direction angles. To reduce the impact of the measurement errors, the authors in \cite{Hu2016} propose a robust DM synthesis method that is capable of achieving an excellent performance of bit error rate  under imperfect direction angle in a single-desired-user scenario. In \cite{Hu2016}, a closed-form projection matrix is derived to force the AN to the null space of steering vector of the desired direction by utilizing the uniform distribution of measurement angle errors. In \cite{Shu2016_2}, the authors propose a robust beamforming scheme for multi-beam DM  broadcasting systems and derive the expressions of different beamformers for various scenarios, in which case only one confidential message stream is broadcasted to multiple desired users.

However, most existing work focuses on robust and non-robust synthesis schemes of DM. Such schemes require perfect knowledge of direction angles or imperfect one with distribution of angle measurement errors at the transmitters. In a practical DM system, it is impossible to obtain perfect knowledge of direction angles or precisely model the distribution of angle measurement errors. This motivates us to develop a low-complexity and robust synthesis scheme needing only the estimated direction angles, without requiring perfect direction angles or the distribution of angle measurement errors under the imperfect case.

In this paper, we consider multiple confidential message stream transmission of using directional modulation in MU-MIMO scenario. We propose a main-lobe-integration (MLI) based leakage synthesis scheme. This method is based on leakage idea in \cite{Tarighat2005, Sadek2007, Shu2011, Shu2016_1} and MLI, and shown robust to angle measurement errors in our simulation. Our main contributions are as follows:

$\bullet$ We propose a totally distinct robust beamfroming method of using the concept of leakage in multi-user MIMO (MU-MIMO) situation, which does not require the statistical knowledge of direction angle measurement error, i.e.,  to estimate the variances of direction error. However, in \cite{Hu2016, Shu2016_2}, the proposed robust methods need to predict the variances of direction estimation errors or even know the distributions of measurement errors in advance.

$\bullet$ The robust method proposed by us simultaneously sends multiple distinct independent  parallel confidential message streams to the associated multi-desired-receivers. However,  those methods in \cite{Hu2016} and \cite{Shu2016_2} transmit only single confidential message stream towards one or more desired users.

$\bullet$  The beamforming vectors for different desired users are individually designed and distinct in this paper. This guarantees that any desired receiver can not intercept confidential messages of others desired user, which further improves the transmit security. However, in \cite{Shu2016_2}, all desired-user channels are combined into one total large virtual channel and only single identical desired beamforming vector is required and devised for all desired users due to broadcasting scenario.

$\bullet$  In the following, AN is viewed as a virtual useful signal for eavesdroppers and we construct the AN projecting matrix by minimizing the leakage of AN power to the main-lobes of the desired directions. This will reduce the effect of AN on desired receivers and maximize the effect of AN on eavesdroppers. In \cite{Shu2016_2}, the AN projection matrix is designed by using the criterion of maximizing receiving-AN-to-signal-noise ratio at undesired receivers.

The rest of the paper is organized as follows. System model is described in Section II. Section III presents the proposed robust synthesis method with desired angle uncertainty under two cases: imperfect and unknown eavesdropper directions. The numerical results are shown and discussed in section IV, and Section V concludes this paper.

\emph{Notations: }throughout the paper, matrices, vectors, and scalars are denoted by letters of bold upper case, bold lower case, and lower case, respectively. Signs $(\cdot)^T$,  $(\cdot)^H$, $(\cdot)^{-1}$ and $\text{tr}(\cdot)$ denote matrix transpose, conjugate transpose, Moore-Penrose inverse and trace, respectively. Operation $(x)^+$ returns zero if $x$ is negative, otherwise $x$ is returned. The notation $\mathbb{E}\{\cdot\}$ refers to the expectation operation. The symbol $\mathbf{I}_N$ denotes the $N \times N$ identity matrix.

Since there are a lot of variables in this paper, we have summarized main variables in Table~\ref{main_var}.
\newcommand{\tabincell}[2]{\begin{tabular}{@{}#1@{}}#2\end{tabular}}
\begin{table}[!ht]
  \centering
  \caption{Summary of main variables}
    \begin{tabular}{p{0.1\columnwidth}<{\centering}p{0.8\columnwidth}}
    \toprule
    $N$ & Number of antennas at the base station(BS)\\
    $K$ & Number of desired users\\
    $M$ & Number of eavesdroppers\\
    $\theta_{d_k}$ & The $k$th desired direction angle\\
    $\theta_{e_m}$ & The $m$th eavesdropping direction angle\\
    $\mathbf{s}$ & Transmit signal vector at the BS\\
    $d_k$ & The $k$th confidential message\\
    $\mathbf{v}_k$ & Beamforming vector of the $k$th confidential message\\
    $\mathbf{T}_{AN}$ & Projection matrix of AN\\
    $\mathbf{z}$ & Random AN vector\\
    $P_s$ & Total transmit power constraint at the BS\\
    $\alpha_1$ & Normalized power factors for confidential messages\\
    $\alpha_2$ & Normalized power factors for AN\\
    $\beta_1$ &  Power allocation of confidential messages\\
    $\beta_2$ &  Power allocation of AN\\
    $\mathbf{h}(\theta)$ & Normalized steering vector along the direction $\theta$\\
    $\varphi_{\theta}(n)$ & Phase difference between the $n$th element in $\mathbf{h}(\theta)$ and the array phase center\\
    $y(\theta)$ & Received signal along the direction $\theta$\\
    $\omega$ & Additive white Gaussian noise(AWGN) with distribution $\mathcal{CN}(0,~\sigma_{\omega}^2)$\\
    $C_k(\theta)$ & Achievable rate of receiving the $k$th useful data along the direction $\theta$\\
    $C_{sec}$ & Secrecy sum-rate\\
    $\hat{\theta}_{d_k}$ & The $k$th estimated desired direction angle\\
    $\hat{\theta}_{e_m}$ & The $m$th estimated eavesdropping direction angle\\
    $\theta_{BW}$ & Beam width between first nulls (BWFN) for a long broadside array\\
    $S$ & Integral interval with single continuous interval or the union of several subintervals\\
    $\mathbf{R}_S$ & Integral result of the matrix $h(\theta)h(\theta)^H$ within the interval $S$, i.e., $\mathbf{R}_S = \int_{S}\mathbf{h}(\theta)\mathbf{h}^H(\theta)\,\text{d}\theta$\\
    \bottomrule
    \end{tabular} \label{main_var}
\end{table}

\section{System Model}
\begin{figure}[!ht]
  \centering
  \includegraphics[width=0.5\textwidth]{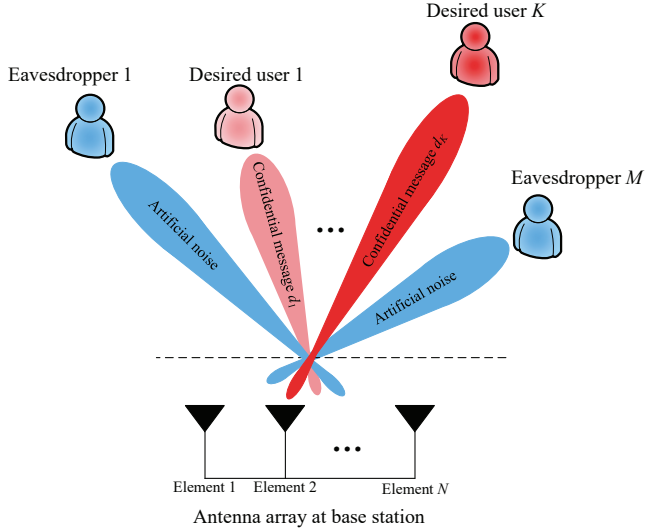}\\
  \caption{Directional modulation MU-MIMO system, where different gray-scale reds denote different confidential message streams for different desired users.}\label{Sys_BlockDiag}
\end{figure}
A schematic diagram of directional modulation MU-MIMO system is shown in Fig.~\ref{Sys_BlockDiag}, where $N$ is the number of elements at the base station (BS). In such a system, we assume that $K$ independent confidential messages $\{d_k\}_{k=1}^K$ are transmitted towards $K$ distinct desired users with direction angles $\{\theta_{d_1}, \theta_{d_2}, \ldots, \theta_{d_K}\}$. Additionally, there are $M$ eavesdroppers with direction angles $\{\theta_{e_1}, \theta_{e_2}, \ldots, \theta_{e_M}\}$. Due to the irregular reflection and refraction in the multipath fading channel, the propagation direction of the signal is unpredictable, while the directional modulation is very sensitive to the arrival direction at desired receiver in the wireless channel. Until now, the extension of directional modulation to multi-path fading channel is a challenging problem. For example, if the blocking object locates at the eavesdropper direction, then it can reflect the AN to the desired receiver. We call the result as the effect of gathering AN. This effect will dramatically degrade the performance of the desired receiver by collecting a large amount of AN from reflecting objects at the undesired directions. Therefore, in this paper, we only consider  the line-of-propagation channels.

The confidential message $d_k$ with $\mathbb{E}\left\{d_k^Hd_{k}\right\}=1$ is sent to the desired receiver $k$, and multiplied by an $N \times 1$ beam-forming vector $\mathbf{v}_k$ before being transmitted through the channel, where $\mathbf{v}_k$ is called the confidential useful vector for desired user $k$ below. Thus, the  transmit signal vector is written as
\begin{equation}\label{Tx_signal_s}
\mathbf{s} =\underbrace{\alpha_1\beta_1\sqrt{P_s} \sum\limits_{k = 1}^K \mathbf{v}_kd_k}_{\text{Confidential messages}}+ \underbrace{\alpha_2\beta_2\sqrt{P_s}\mathbf{T}_{AN}\mathbf{z}}_{\text{AN}},
\end{equation}
where $P_s$ is the total transmit power constraint at BS, $\beta_1$  and $\beta_2$ are the power allocation between confidential messages and AN such that
\begin{equation}\label{beta_square_sum}
\beta^2_1+\beta^2_2 =1,
\end{equation}
$\alpha_1$ and $\alpha_2$ are the normalized power factors for confidential messages and AN such that
\begin{equation}\label{normalized_AN}
\alpha^2_2 \mathbb{E}\left\{\text{tr}\left[\mathbf{T}_{AN}\mathbf{z}\mathbf{z}^H\mathbf{T}_{AN}^H\right]\right\}=1,
\end{equation}
and $\mathbf{T}_{AN}$ is the projection matrix of forcing AN to the eavesdropping directions. The transmit signal vector in (\ref{Tx_signal_s}) satisfies the following power constraint
\begin{align}\label{normalized_vd}
\alpha^2_1 \mathbb{E}\left\{\sum\limits_{k = 1}^K\sum\limits_{k' = 1}^K \mathbf{v}_{k'}^H\mathbf{v}_kd_kd_{k'}^H\right\}=1.
\end{align}
If $\mathbb{E}\left\{d_kd_{k}^H\right\}=1$, $\mathbf{v}_k^H \mathbf{v}_k=1$, and $\mathbb{E}\left\{\mathbf{z}\mathbf{z}^H\right\}=\frac{1}{N-K}\mathbf{I}_{N-K}$, (\ref{normalized_AN}) and (\ref{normalized_vd}) can be further simplified as
\begin{equation}\label{alpha_2}
\alpha^2_2 \text{tr}\left[\mathbf{T}_{AN}\mathbf{T}_{AN}^H\right]=N-K,
\end{equation}
and
\begin{equation}\label{alpha_1}
\alpha^2_1=\frac{1}{K}.
\end{equation}

The $N\times1$ vector $\mathbf{s}$ passes through the LoS channel, the received signal along direction $\theta$ is given by
\begin{equation}\label{Rx_signal_y}
y(\theta) = \mathbf{h}^H(\theta)\mathbf{s}+\omega,
\end{equation}
where $\omega$ is the additive white Gaussian noise (AWGN) with distribution $\mathcal {C}\mathcal {N}(0, \sigma_\omega^2)$, and
\begin{equation}\label{h_theta}
\mathbf{h}(\theta)=\frac{1}{\sqrt{N}}\left[\underbrace{e^{j2\pi\varphi_{\theta}(1)}}_{h_1(\theta)}, \cdots, \underbrace{e^{j2\pi\varphi_{\theta}(n)}}_{h_n(\theta)}, \cdots, \underbrace{e^{j2\pi\varphi_{\theta}(N)}}_{h_N(\theta)}\right]^T
\end{equation}
is the normalized steering vector along the direction $\theta$ with function $\varphi_{\theta}(n)$ defined by
\begin{equation}\label{var_phi}
\varphi_{\theta}(n)\triangleq\tfrac{(n-(N+1)/2)d\cos\theta}{\lambda}, (n=1,2,\ldots,N),
\end{equation}
where $d$ denotes the spacing between two adjacent elements of transmit antenna array and $\lambda$ is the wavelength of transmit carrier. According to (\ref{Rx_signal_y}), the received signal at the $k$th desired user is given by
\newcounter{mytempeqncnt}

\begin{figure*}[!t]
\normalsize
\setcounter{mytempeqncnt}{\value{equation}}
\begin{equation}\label{Ck_theta_dk}
\begin{split}
C_k(\theta_{d_k})
 & \triangleq I(y(\theta_{d_k});[d_k,\theta_{d_k}])\\
 & = \log_2(1+\frac{\alpha_1^2\beta_1^2P_s \mathbf{h}^H(\theta_{d_k})\mathbf{v}_k\mathbf{v}_k^H\mathbf{h}(\theta_{d_k})}{\sigma^2_{d_k}+\sum\limits_{i = 1, i \neq k}^K\alpha_1^2\beta_1^2P_s \mathbf{h}^H(\theta_{d_k})\mathbf{v}_i\mathbf{v}_i^H\mathbf{h}(\theta_{d_k})+\alpha_2^2\beta_2^2P_s \mathbf{h}^H(\theta_{d_k})\mathbf{T}_{AN}\mathbf{T}_{AN}^H\mathbf{h}(\theta_{d_k})}).
\end{split}
 \tag{13}
\end{equation}
\begin{equation}\label{Ck_theta_em}
\begin{split}
C_k(\theta_{e_m})
 & \triangleq I(y(\theta_{e_m});[d_k,\theta_{e_m}])\nonumber\\
 & = \log_2(1+\frac{\alpha_1^2\beta_1^2P_s \mathbf{h}^H(\theta_{e_m})\mathbf{v}_k\mathbf{v}_k^H\mathbf{h}(\theta_{e_m})}{\sigma^2_{e_m}+\sum\limits_{i = 1, i \neq k}^K\alpha_1^2\beta_1^2P_s \mathbf{h}^H(\theta_{e_m})\mathbf{v}_i\mathbf{v}_i^H\mathbf{h}(\theta_{e_m})+\alpha_2^2\beta_2^2P_s \mathbf{h}^H(\theta_{e_m})\mathbf{T}_{AN}\mathbf{T}_{AN}^H\mathbf{h}(\theta_{e_m})}).
\end{split}
 \tag{14}
\end{equation}
\setcounter{equation}{\value{mytempeqncnt}}
\hrulefill
\vspace*{4pt}
\end{figure*}
\begin{align}\label{Rx_signal_ydk}
& y(\theta_{d_k}) = \mathbf{h}^H(\theta_{d_k})\mathbf{s}+\omega_{d_k}\nonumber\\
& = \underbrace{\alpha_1\beta_1\sqrt{P_s} \mathbf{h}^H(\theta_{d_k})\mathbf{v}_kd_k}_{\text{Useful data}}+\underbrace{\alpha_1\beta_1\sqrt{P_s} \mathbf{h}^H(\theta_{d_k}) \sum\limits_{i = 1, i \neq k}^K \mathbf{v}_id_i}_{\text{Interference from other users}}\nonumber\\
& + \underbrace{\alpha_2\beta_2\sqrt{P_s} \mathbf{h}^H(\theta_{d_k})\mathbf{T}_{AN}\mathbf{z}}_{\text{AN}}+\underbrace{\omega_{d_k}}_{\text{AWGN}},
\end{align}
where the first term of the above expression is the useful received signal for user $k$, the second one is the multi-user interference from other users, the third one is the AN, and the last one is the AWGN with distribution $\mathcal {C}\mathcal {N}(0, \sigma^2_{d_k})$ at receiver.
Similarly, the received signal at eavesdropper $m$ is
\begin{align}\label{Rx_signal_yem}
y(\theta_{e_m}) & = \mathbf{h}^H(\theta_{e_m})\mathbf{s}+\omega_{e_m}\nonumber\\
& = \underbrace{\alpha_1\beta_1\sqrt{P_s} \mathbf{h}^H(\theta_{e_m}) \sum\limits_{k = 1}^K \mathbf{v}_kd_k}_{\text{Confidential messages}}\nonumber\\
& + \underbrace{\alpha_2\beta_2\sqrt{P_s} \mathbf{h}^H(\theta_{e_m})\mathbf{T}_{AN}\mathbf{z}}_{\text{AN}} +\underbrace{\omega_{e_m}}_{\text{AWGN}},
\end{align}
where the first term of the above expression is composed of the confidential messages intercepted by eavesdropper $m$, the second one is the AN to disturb eavesdropper $m$, and the last one is the AWGN with distribution $\mathcal {C}\mathcal {N}(0, \sigma^2_{e_m})$.


To evaluate the security performance for multi-user scenario in this paper, the secrecy sum-rate is adopted and can be defined as the sum of difference in available rate receiving useful data between secure transmission channel and eavesdropper channel, \cite{Oggier2011}, \cite{Li2014},
\begin{equation}\label{Csec}
C_{sec}=\sum\limits_{k=1}^K\left[C_k(\theta_{d_k})-\max\limits_mC_k(\theta_{e_m})\right]^+,
\end{equation}
where $C_k(\theta_{d_k})$ and $C_k(\theta_{e_m})$ shown in (\ref{Ck_theta_dk}) and (\ref{Ck_theta_em}) are the achievable rates of receiving useful data $d_k$ along the $k$th desired direction $\theta_{d_k}$ and the $m$th eavesdropper direction $\theta_{e_m}$, respectively. Function $I(y;[d,\theta])$ denotes the mutual information along direction $\theta$ between the input $d$ and the output $y$.

The secrecy sum-rate in (\ref{Csec}) is one of the most important metrics for assessing the performance of the DM system. Under the multi-user scenario, the expression of the secrecy sum-rate in (\ref{Csec}) contains multiple variables, i.e., $R_k(\theta_{d_k})$ and $R_k(\theta_{e_m})$. In addition, $R_k(\theta_{d_k})$ and $R_k(\theta_{e_m})$ are both $K+1$ coupled variables with $\{\mathbf{v}_{d_k}\}_{k=1}^K$ and $\mathbf{T}_{AN}$ as shown in (\ref{Ck_theta_dk}) and (\ref{Ck_theta_em}), respectively. Apparently, it is challenging to solve the $K+1$ coupled optimization problems. Maximizing the secrecy sum-rate in (\ref{Csec}) directly is an NP-hard problem with exponential complexity. Maybe, with a high-computational amount, only suboptimal solution is achieved. On the other hand, minimizing the bit error rate (BER), which is another important metric for evaluating the DM system, of a selected user will depend heavily on other users' beamforming vectors and direction angles. The coupled property requires the iteration operation. Additionally, it is not easy to guarantee its convergence. To address the above, we propose a main-lobe-integration based leakage beamforming method, which can provide an approximate closed-form solution with low-complexity and high performance. More importantly, due to main-lobe-integration, it is also robust to direction angle measurement errors.

\section{Proposed Robust Synthesis Method with Desired Angle Uncertainty }
\begin{figure}[!ht]
  \centering
  \includegraphics[width=0.5\textwidth]{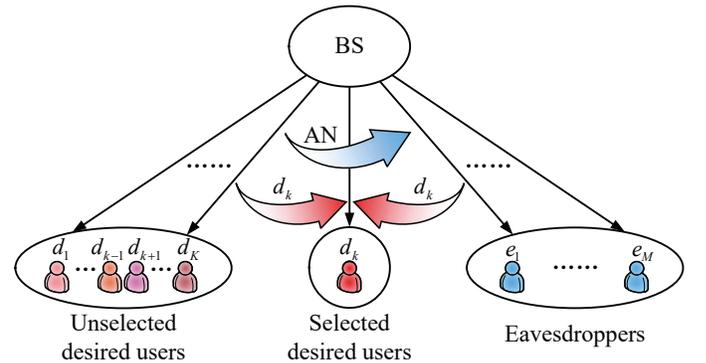}\\
  \caption{Schematic diagram of the proposed scheme.}\label{Leak_AN_Idea}
\end{figure}

In Fig.~\ref{Leak_AN_Idea}, we sketch the basic idea of the proposed robust precoding method based on MLI and leakage. As shown In Fig.~\ref{Leak_AN_Idea}, if desired user $k$ is chosen as the current desired user, the corresponding confidential useful beamforming vector is given by minimizing its useful signal power leakage to the remaining desired users and all eavesdroppers. Confidential beamforming vector corresponding to each desired user is designed individually in order to safeguard each desired user privacy. The AN projection matrix $\mathbf{T}_{AN}$ is optimized by maximizing the AN power leakage to all eavesdroppers. In other words, the influence of AN  on all desired users is minimized. Here, $\mathbf{T}_{AN}$ is constructed in the all-in-one way not individually. In the following, we mainly consider the imperfect desired direction, where imperfect means that there exists with measurement errors on the measured desired direction angles. Eavesdropper direction angles fall into two categories: imperfect (See Subsection III.A) and unknown (See Subsection III.B). In the first scenario, both  desired and eavesdropper directions are imperfect. In other words, there usually exists errors in the measured desired and eavesdropper directions.  In the second scenario, eavesdropper directions are unknown while the measured desired directions are imperfect. In the two situations, we accordingly show how to design the AN projection matrix $\mathbf{T}_{AN}$ and confidential useful beamforming vectors $\mathbf{v}_{d_k}$.

\subsection{Imperfect desired and eavesdropper directions}
In practice, the desired and eavesdropper directions are unknown. In this case, the BS can estimate both the values of desired and eavesdropper direction angles with traditional spatial spectrum estimation such as MUSIC, Capon and ESPRIT \cite{Gross2005}. Due to the effect of channel noise, there always exist errors in the estimated direction angles. Given the estimated direction angles $\hat{\theta}_{d_k}$ and $\hat{\theta}_{e_m}$ associated with desired user $k$ and eavesdropper $m$, we define their main-lobe intervals  as
\begin{equation}\label{S_dk}
S_{d_k}=\left[\hat{\theta}_{d_k}-\frac{\theta_{BW}}{2},~ \hat{\theta}_{d_k}+\frac{\theta_{BW}}{2}\right],
\setcounter{equation}{15}
\end{equation}
and
\begin{equation}\label{S_em}
S_{e_m}=\left[\hat{\theta}_{e_m}-\frac{\theta_{BW}}{2},~\hat{\theta}_{e_m}+\frac{\theta_{BW}}{2}\right],
\end{equation}
where
\begin{equation}\label{theta_BW}
\theta_{BW}=\frac{2\lambda}{Nd}
\end{equation}
is the beam width between first nulls (BWFN) for a long broadside array \cite{Kraus2006}. Therefore, the overall direction angle intervals of all main-lobes of desired users and eavesdroppers are
\begin{equation}\label{S_d}
S_d=\bigcup\limits_{k=1}^K\left[\hat{\theta}_{d_k}-\frac{\theta_{BW}}{2}, \hat{\theta}_{d_k}+\frac{\theta_{BW}}{2}\right],~\bar{S}_d=\left[0,~\pi\right]\backslash S_d,
\end{equation}
and
\begin{equation}\label{S_e}
S_e=\bigcup\limits_{m=1}^M\left[\hat{\theta}_{e_m}-\frac{\theta_{BW}}{2},~\hat{\theta}_{e_m}+\frac{\theta_{BW}}{2}\right],
\end{equation}
respectively. Based on the above discussion, the average power of confidential message stream $d_k$ sent to  desired user $k$ is
\begin{align}\label{P_dk_U_eave_known}
P_{d_k,U}&= \mathbb{E}\left\{\int_{S_{d_k}}\alpha_1^2\beta_1^2P_sd_k^H\mathbf{v}_{d_k}^H\mathbf{h}(\theta)\mathbf{h}^H(\theta)\mathbf{v}_{d_k}d_k\,d\theta\right\} \nonumber\\
&= \int_{S_{d_k}}\alpha_1^2\beta_1^2P_s\mathbf{v}_{d_k}^H\mathbf{h}(\theta)\mathbf{h}^H(\theta)\mathbf{v}_{d_k}\,d\theta.
\end{align}
The remaining power leakage  of confidential message stream $d_k$ to other desired users and all eavesdroppers is represented as
\begin{align}\label{P_dk_L_eave_known0}
& P_{d_k,L}\nonumber\\
& = \sum\limits_{i=1,i \neq k}^K \mathbb{E}\left\{\int_{S_{d_i}}\alpha_1^2\beta_1^2P_sd_k^H\mathbf{v}_{d_k}^H\mathbf{h}(\theta)\mathbf{h}^H(\theta)\mathbf{v}_{d_k}d_k\,d\theta\right\}\nonumber\\
& + \sum\limits_{m=1}^M\mathbb{E}\left\{\int_{S_{e_m}}\alpha_1^2\beta_1^2P_sd_k^H\mathbf{v}_{d_k}^H\mathbf{h}(\theta)\mathbf{h}^H(\theta)\mathbf{v}_{d_k}d_k\,d\theta\right\},
\end{align}
which can be simplified as
\begin{align}\label{P_dk_L_eave_known1}
P_{d_k,L} & = \sum\limits_{i=1,i \neq k}^K \int_{S_{d_i}}\alpha_1^2\beta_1^2P_s\mathbf{v}_{d_k}^H\mathbf{h}(\theta)\mathbf{h}^H(\theta)\mathbf{v}_{d_k}\,d\theta\nonumber\\
& + \sum\limits_{m=1}^M\int_{S_{e_m}}\alpha_1^2\beta_1^2P_s\mathbf{v}_{d_k}^H\mathbf{h}(\theta)\mathbf{h}^H(\theta)\mathbf{v}_{d_k}\,d\theta,
\end{align}
where the first and second terms are the leakage powers to other desired users and all eavesdroppers, respectively. We define the corresponding MLI-SLNR as follows
\begin{equation}\label{SLNR_dk_eave_known_1}
\text{MLI-SLNR}(\mathbf{v}_{d_k}) = \frac{ P_{d_k,U}}{\int_{S_{d_k}}\sigma^2_{d_k}\,d\theta+P_{d_k,L}}.
\end{equation}
which, according to Appendix A, can be simplified as
\begin{align}\label{SLNR_dk_eave_known_2}
\text{MLI-SLNR}&(\mathbf{v}_{d_k})=\nonumber\\
&\frac{\mathbf{v}_{d_k}^H\mathbf{R}_{S_{d_k}}\mathbf{v}_{d_k}}{\mathbf{v}_{d_k}^H\left(\frac{\sigma_{d_k}^2\theta_{BW}}{\alpha_1^2\beta_1^2P_s}\mathbf{I}_N+\mathbf{R}_{S_d\backslash S_{d_k}}+\mathbf{R}_{S_e}\right)\mathbf{v}_{d_k}},
\end{align}
where matrix $\mathbf{R}_S$ is defined as
\begin{equation}\label{R_S}
\mathbf{R}_S = \int_{S}\mathbf{h}(\theta)\mathbf{h}^H(\theta)\,d\theta,
\end{equation}
where $S$ is the integral interval with single continuous interval or the union of several subintervals and each element of $\mathbf{R}_S$ is the definite integration of the corresponding element in the $N \times N$ matrix $\mathbf{h}(\theta)\mathbf{h}^H(\theta)$ , which is a function of $\theta$, over the interval $S$.
The detailed derivation of matrix $\mathbf{R}_S$ is given in Appendix B. Maximizing the MLI-SLNR in (\ref{SLNR_dk_eave_known_2}) by using the generalized Rayleigh-Ritz theorem in \cite{Horn1987} yields that the optimal $\mathbf{v}_{d_k}$ is the normalized eigenvector corresponding to the largest eigenvalue of
\begin{equation}\label{vdk_eave_known}
\left[\frac{\sigma_{d_k}^2\theta_{BW}}{\alpha_1^2\beta_1^2P_s}\mathbf{I}_N+\mathbf{R}_{S_d\backslash S_{d_k}}+\mathbf{R}_{S_e}\right]^{-1}\mathbf{R}_{S_{d_k}}.
\end{equation}

Until now we complete the design of $\mathbf{v}_{d_k}$. Below, we similarly construct the projection matrix $\mathbf{T}_{AN}$ of AN. The basic idea is to project less power of AN onto the subspace spanned by all desired steering vectors and more onto its null space. Here, we view the AN as a useful signal.
The average AN power sent to all main-lobes of all eavesdropper directions is as follows
\begin{align}\label{P_AN_U_eave_known_1}
&P_{AN,U}\nonumber\\
&=\mathbb{E}\left\{\int_{S_e}\alpha_2^2\beta_2^2P_s\text{tr}\left[\mathbf{h}^H(\theta)\mathbf{T}_{AN}\mathbf{z}\mathbf{z}^H\mathbf{T}_{AN}^H\mathbf{h}(\theta)\right]\,d\theta\right\}\nonumber\\
&=\int_{S_e}\alpha_2^2\beta_2^2P_s\text{tr}\left[\mathbf{h}^H(\theta)\mathbf{T}_{AN}\mathbb{E}\left\{\mathbf{z}\mathbf{z}^H\right\}\mathbf{T}_{AN}^H\mathbf{h}(\theta)\right]\,d\theta,
\end{align}
Given $\mathbb{E}\left\{\mathbf{z}\mathbf{z}^H\right\}=\frac{1}{N-K}\mathbf{I}_{N-K}$ and $\text{tr}\left(\mathbf{A}\mathbf{B}\right)=\text{tr}\left(\mathbf{B}\mathbf{A}\right)$, the above equation is reduced to
\begin{align}\label{P_AN_U_eave_known_2}
P_{AN,U}
&=\frac{\alpha_2^2\beta_2^2P_s}{N-K}\text{tr}\left[\mathbf{T}_{AN}^H\mathbf{R}_{S_e}\mathbf{T}_{AN}\right].
\end{align}
The average leakage power of AN to main-lobes of all desired directions is given by
\begin{align}\label{P_AN_L_eave_known_1}
&P_{AN,L}\nonumber\\
&=\mathbb{E}\left\{\int_{S_d}\alpha_2^2\beta_2^2P_s\text{tr}\left[\mathbf{z}^H\mathbf{T}_{AN}^H\mathbf{h}(\theta)\mathbf{h}^H(\theta)\mathbf{T}_{AN}\mathbf{z}\right]\,d\theta\right\}.
\end{align}
Similar to (\ref{P_AN_U_eave_known_2}), we have
\begin{align}\label{P_AN_L_eave_known_2}
P_{AN,L}&=\frac{\alpha_2^2\beta_2^2P_s}{N-K}\text{tr}\left[\mathbf{T}_{AN}^H\mathbf{R}_{S_d}\mathbf{T}_{AN}\right].
\end{align}
Combining the above two expressions and using the definition of SLNR, we have the MLI-SLNR of AN as
\begin{align}\label{SLNR_AN_eave_known}
\text{MLI-SLNR}(\mathbf{T}_{AN})
& = \frac{P_{AN,U}}{\int_{S_e}\sigma^2_e\,d\theta+P_{AN,L}}\nonumber\\
& = \frac{\text{tr}\left[\mathbf{T}_{AN}^H\mathbf{R}_{S_e}\mathbf{T}_{AN}\right]}{\frac{(N-K)M\theta_{BW}\sigma^2_e}{\alpha_2^2\beta_2^2P_s}+\text{tr}\left[\mathbf{T}_{AN}^H\mathbf{R}_{S_d}\mathbf{T}_{AN}\right]},
\end{align}
where
\begin{equation}\label{sigma_e}
\sigma_e^2=\frac{1}{M}\sum\limits_{m=1}^M\sigma_{e_m}^2.
\end{equation}
As a result, the optimal $\mathbf{T}_{AN}$ when maximizing the MLI-SLNR in (\ref{SLNR_AN_eave_known}) is composed of the $N-K$ normalized eigenvectors corresponding to the $N-K$ largest eigenvalues of matrix
\begin{equation}\label{T_AN_eave_known}
\left[\frac{(N-K)M\theta_{BW}\sigma^2_e}{\alpha_2^2\beta_2^2P_s}\mathbf{I}_N+\mathbf{R}_{S_d}\right]^{-1}\mathbf{R}_{S_e}.
\end{equation}

\subsection{Unknown eavesdropper directions}
In the following, we consider a more practical scenario, where the BS doesn't know the exact or estimated direction values of eavesdroppers. In such a situation, all the remaining angle region excluding the union of main-lobes of all desired directions, which is actually the complementary set $\bar{S}_{d}$ of $S_{d}$ , will be viewed as the potential eavesdropper directions. For desired user $k$, the main-lobe profile of all eavesdropper and the remaining desired directions are virtually viewed as a potential set of intercepting directions, defined as the complementary set $\bar{S}_{d_k}$ of its main-lobe region $S_{d_k}$, i.e.,
\begin{align}\label{S_dk_bar}
~\bar{S}_ {d_k}&=\left[0,~\pi\right]\backslash S_{d_k}\nonumber\\&=\left[0,~\hat{\theta}_{d_k}-\frac{\theta_{BW}}{2}\right]\bigcup\left[\hat{\theta}_{d_k}+\frac{\theta_{BW}}{2},~\pi\right].
\end{align}
Note that the useful part $P'_{d_k,U}$ of transmit signal power of desired user $k$ has the same expression as $P_{d_k,U}$ in (\ref{P_dk_U_eave_known}). The power leakage of confidential message stream $d_k$ transmitted by BS to the set $~\bar{S}_{d_k}$ of all potential eavesdropper directions is represented as
\begin{align}\label{P_dk_U_eave_unknown}
P'_{d_k,L} & = \mathbb{E}\left\{\int_{\bar{S}_{d_k}}\alpha_1^2\beta_1^2P_sd_k^H\mathbf{v}_{d_k}^H\mathbf{h}(\theta)\mathbf{h}^H(\theta)\mathbf{v}_{d_k}d_k\,d\theta\right\}\nonumber\\
& = \int_{\bar{S}_{d_k}}\alpha_1^2\beta_1^2P_s\mathbf{v}_{d_k}^H\mathbf{h}(\theta)\mathbf{h}^H(\theta)\mathbf{v}_{d_k}\,d\theta.
\end{align}
Similar to (\ref{SLNR_dk_eave_known_1}), we can readily obtain the MLI-SLNR expression corresponding to the $k$th desired user as follows
\begin{align}\label{SLNR_dk_eave_unknown}
\text{MLI-SLNR}'(\mathbf{v}_{d_k})
& = \frac{ P'_{d_k,U}}{\int_{S_{d_k}}\sigma^2_{d_k}\,d\theta+P'_{d_k,L}}\nonumber\\
& = \frac{\mathbf{v}_{d_k}^H\mathbf{R}_{S_{d_k}}\mathbf{v}_{d_k}}{\mathbf{v}_{d_k}^H\left(\frac{\sigma_{d_k}^2\theta_{BW}}{\alpha_1^2\beta_1^2P_s}\mathbf{I}_N+\mathbf{R}_{\bar{S}_{d_k}}\right)\mathbf{v}_{d_k}}.
\end{align}
The optimal $\mathbf{v}_{d_k}$ to maximize the MLI-SLNR in (\ref{SLNR_dk_eave_unknown}) is the generalized eigenvector corresponding to the largest normalized eigenvalue of
\begin{equation}\label{vdk_eave_unknown}
\left[\frac{\sigma_{d_k}^2\theta_{BW}}{\alpha_1^2\beta_1^2P_s}\mathbf{I}_N+\mathbf{R}_{\bar{S}_{d_k}}\right]^{-1}\mathbf{R}_{S_{d_k}}.
\end{equation}

For the design of $\mathbf{T}_{AN}$ with unknown directions of eavesdroppers, the angle range of all potential eavesdropping directions is $\bar{S}_d$, the average transmit AN power in $\bar{S}_d$ is
\begin{align}\label{P_AN_U_eave_unknown}
&P'_{AN,U}\nonumber\\
&=\mathbb{E}\left\{\int_{\bar{S}_d}\alpha_2^2\beta_2^2P_s\text{tr}\left[\mathbf{h}^H(\theta)\mathbf{T}_{AN}\mathbf{z}\mathbf{z}^H\mathbf{T}_{AN}^H\mathbf{h}(\theta)\right]\,d\theta\right\}\nonumber\\
&=\frac{\alpha_2^2\beta_2^2P_s}{N-K}\text{tr}\left[\mathbf{T}_{AN}^H\mathbf{R}_{\bar{S}_d}\mathbf{T}_{AN}\right],
\end{align}
and the average leakage AN power $P'_{AN,L}$ to main-lobes of all desired directions is the same as (\ref{P_AN_L_eave_known_2}). Based on the definition of SLNR, we have the MLI-SLNR expression of AN as
\begin{align}\label{SLNR_AN_eave_unknown}
&\text{MLI-SLNR}'(\mathbf{T}_{AN})\nonumber\\
& = \frac{P'_{AN,U}}{\int_{\bar{S}_d}\sigma^2_e\,d\theta+P'_{AN,L}}\nonumber\\
& = \frac{\text{tr}\left[\mathbf{T}_{AN}^H\mathbf{R}_{\bar{S}_d}\mathbf{T}_{AN}\right]}{\frac{(N-K)(~\pi-K\theta_{BW})\sigma^2_e}{\alpha_2^2\beta_2^2P_s}+\text{tr}\left[\mathbf{T}_{AN}^H\mathbf{R}_{S_d}\mathbf{T}_{AN}\right]}.
\end{align}
Similar to (\ref{T_AN_eave_known}), via maximizing the MLI-SLNR in (\ref{SLNR_AN_eave_unknown}), the optimal $\mathbf{T}_{AN}$ is composed of the $N-K$ normalized eigenvectors corresponding to the $N-K$ largest eigenvalues of matrix
\begin{equation}\label{T_AN_eave_unknown}
\left[\frac{(N-K)(~\pi-K\theta_{BW})\sigma^2_e}{\alpha_2^2\beta_2^2P_s}\mathbf{I}_N+\mathbf{R}_{S_d}\right]^{-1}\mathbf{R}_{\bar{S}_d}.
\end{equation}

\section{Simulation and Discussion}
To evaluate the bit error rate (BER) and secrecy-sum-rate (SSR) performance of the proposed robust method,  quadrature phase shift keying (QPSK) is chosen, and main simulation parameters are listed in Table~\ref{main_para}.
\begin{table}[!ht]
  \centering
  \caption{Main simulation parameters and their values}
    \begin{tabular}{|p{0.35\columnwidth}<{\centering}|p{0.25\columnwidth}<{\centering}|}
    \hline
    \textbf{Parameter} & \textbf{Value}  \\
    \hline
    $d$ & $\lambda/2$\\
    \hline
    $N$ & 16\\
    \hline
    $K$ & 2\\
    \hline
    $\{\theta_{d_k}\}_{k=1}^K$ & $\{60^\circ, 120^\circ\}$\\
    \hline
    $M$ & 3\\
    \hline
    $\{\theta_{e_m}\}_{m=1}^M$ & $\{30^\circ, 90^\circ, 150^\circ\}$\\
    \hline
    $P_s$ & $1$\\
    \hline
    $\beta_1$ & $\sqrt{0.9}$\\
    \hline
    $\beta_2$ & $\sqrt{0.1}$\\
    \hline
    $\Delta\theta_{\text{max}}$ & $5^\circ$\\
    \hline
    \end{tabular} \label{main_para}
\end{table}

Here, signal-to-noise ratio is defined as $\text{SNR}=10\log_{10}(\alpha_1^2\beta_1^2P_s/\sigma_\omega^2)$, where $\sigma_{d_k}^2=\sigma_{e_m}^2=\sigma_\omega^2, \forall k\in\{1, 2\}, \forall m\in\{1, 2, 3\}$. All measurement errors of desired and eavesdropper direction angles are assumed to be independently uniform distributed over the interval $[-\Delta\theta_{\text{max}}, \Delta\theta_{\text{max}}]$. The orthogonal projection (OP) method in \cite{Ding2015_2} and the conventional leakage-based method in \cite{Sadek2007} are adopted as performance references, respectively. All the remaining angle region excluding the union of main-lobes of all desired directions will be viewed as the potential eavesdropper directions when eavesdroppers' information are unable to be available.

\begin{figure}[!t]
\centering
\subfigure[$\theta_{d_1}=60^\circ$]{
\label{BER_eave_known_sub1}
\includegraphics[width=0.5\textwidth]{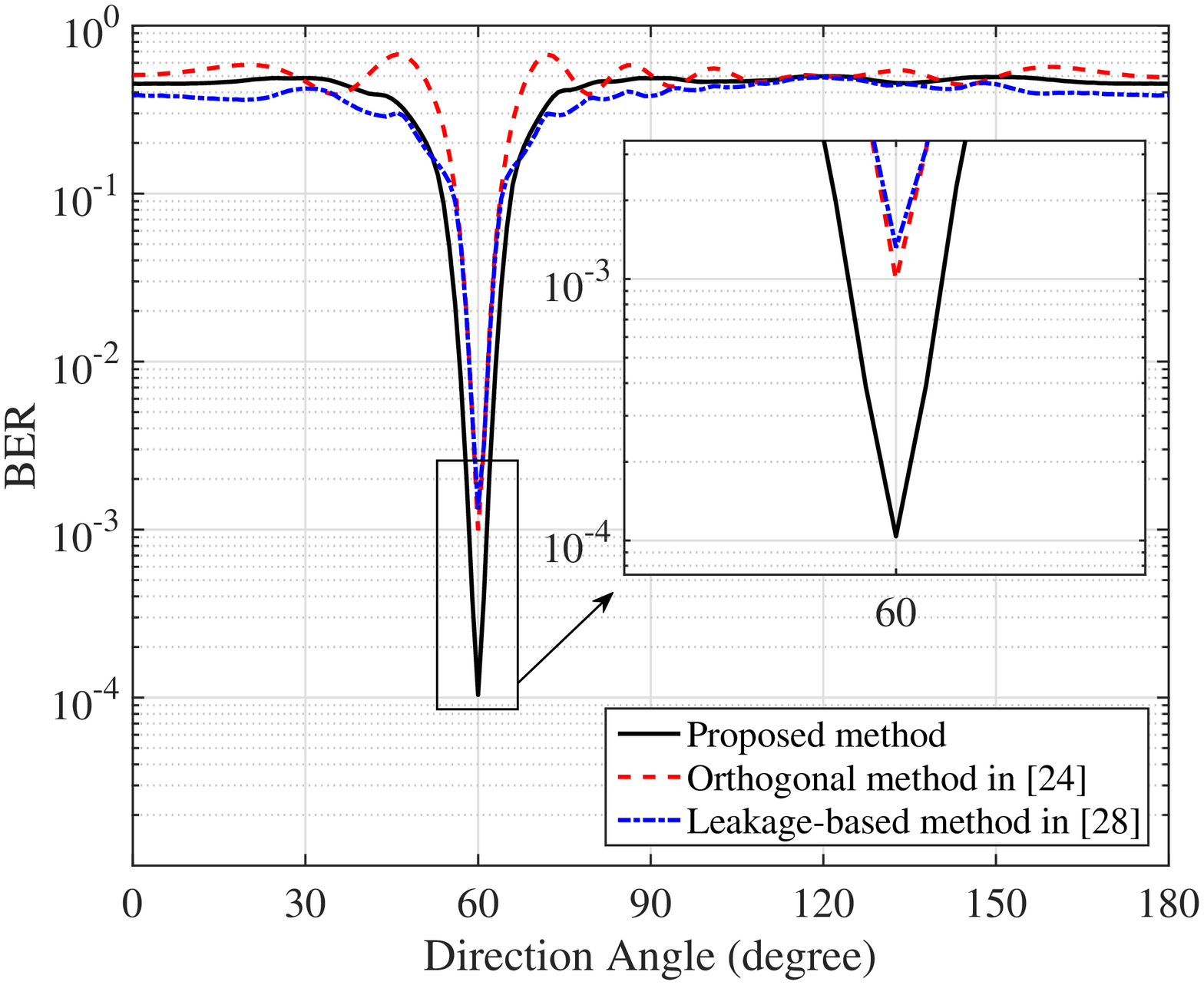}}
\hspace{1in}
\subfigure[$\theta_{d_2}=120^\circ$]{
\label{BER_eave_known_sub2}
\includegraphics[width=0.5\textwidth]{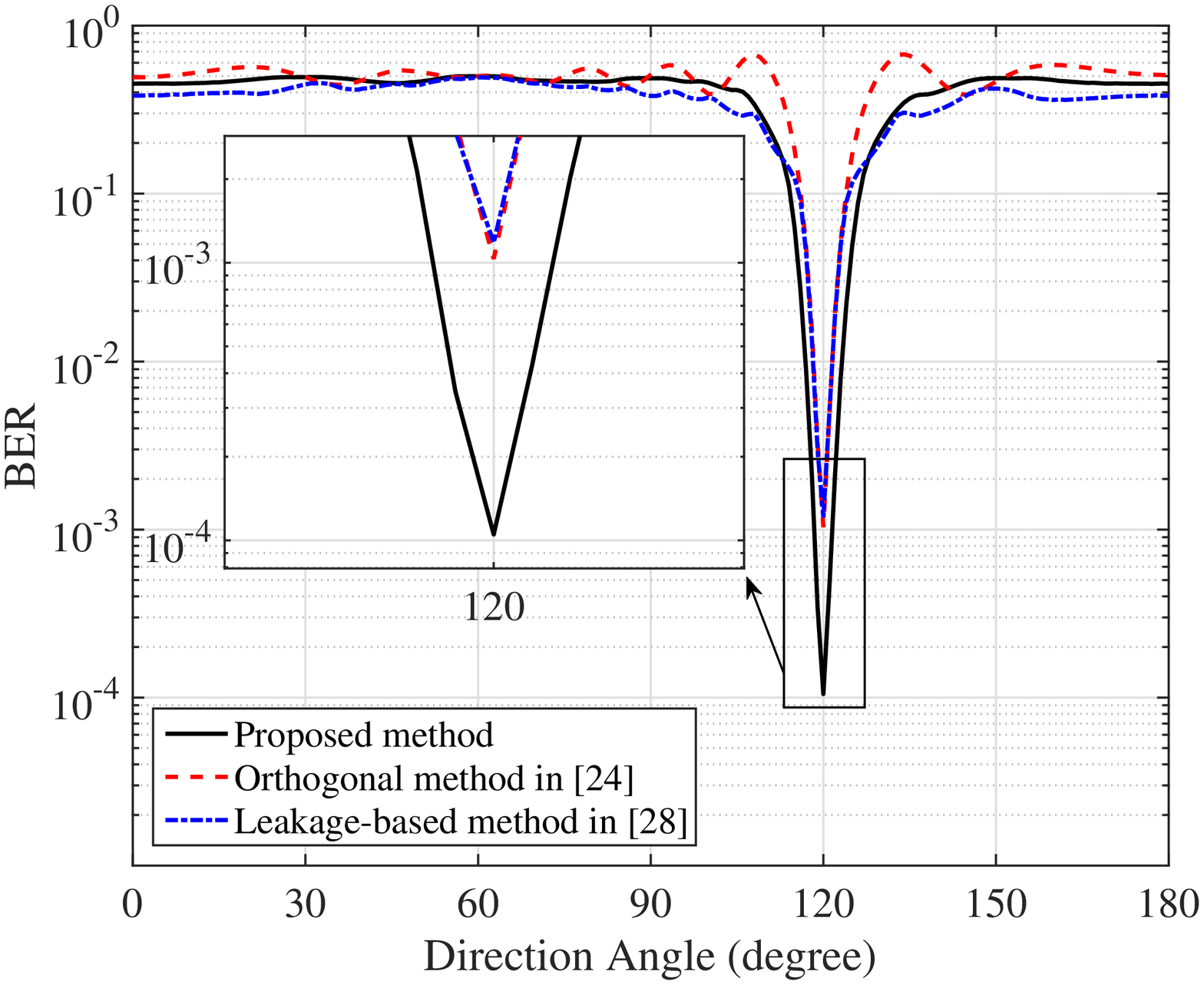}}
\caption{BER versus direction angle with imperfect desired and eavesdropper directions (SNR=14dB).}
\label{BER_eave_known}
\end{figure}

\begin{figure}[!t]
\centering
\subfigure[$\theta_{d_1}=60^\circ$]{
\label{BER_eave_unknown_sub1}
\includegraphics[width=0.5\textwidth]{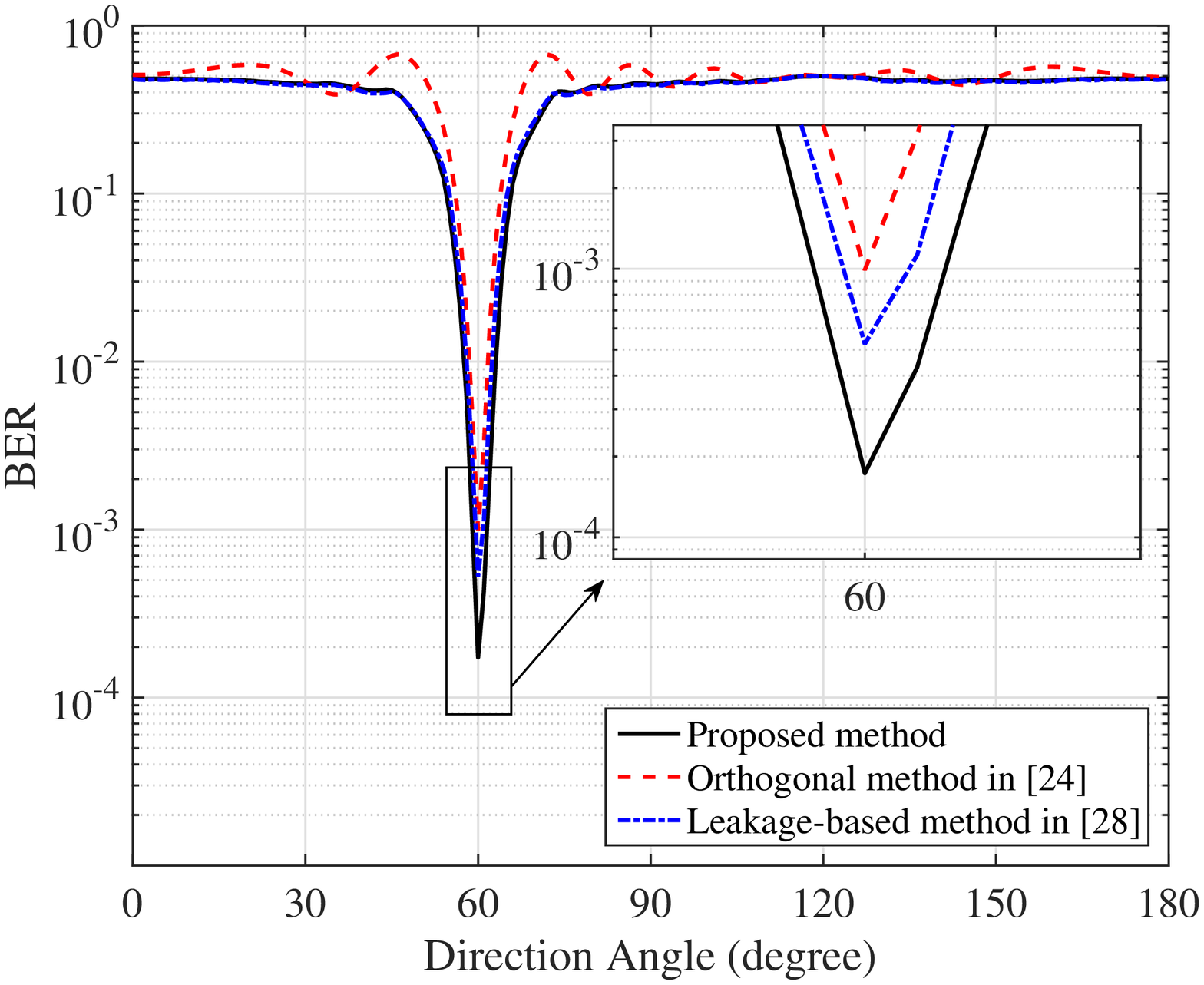}}
\hspace{1in}
\subfigure[$\theta_{d_2}=120^\circ$]{
\label{BER_eave_unknown_sub2}
\includegraphics[width=0.5\textwidth]{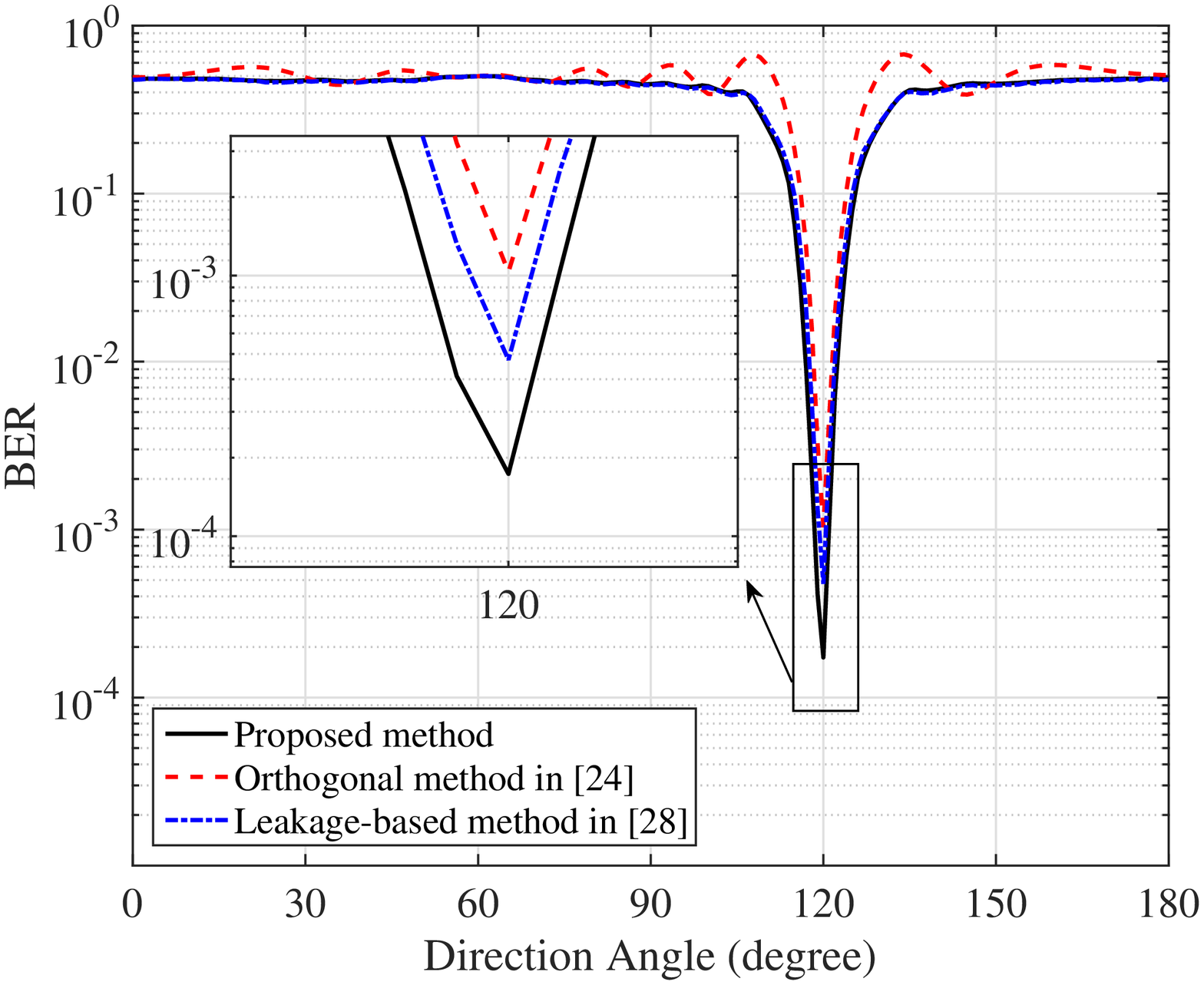}}
\caption{BER versus direction angle with unknown eavesdropper directions (SNR=14dB).}
\label{BER_eave_unknown}
\end{figure}

Given the estimated desired and eavesdropper directions, Fig.~\ref{BER_eave_known}~(a) and (b) illustrate the curves of BER versus direction angle of the proposed robust method in Section III-A, the OP method in \cite{Ding2015_2} and the conventional leakage-based method in \cite{Sadek2007} for the first and second desired receivers, respectively. In the Fig.~\ref{BER_eave_known}~(a), it can be observed that our proposed method at the desired direction $60^\circ$ shows about an order-of-magnitude improvement over that of the OP method and the conventional leakage-based method. The lowest BER values of the OP method and the conventional leakage-based method are approximately at the same level. As the receiver direction angle steers away from the desired direction, the BER performance of both methods degrade rapidly. Compared to the remaining two methods, the BER curve of our proposed method fluctuates less outside the main-lobe of the desired direction. For the second desired user with direction $\theta_{d_2}=120^\circ$ in Fig.~\ref{BER_eave_known}~(b), the performance trend is similar to the first desired user.

Given the estimated desired directions and under unknown eavesdropper directions, Fig.~\ref{BER_eave_unknown}~(a) and (b) plot the curves of BER versus direction angle of the proposed robust method in Section III-B, the OP method in \cite{Ding2015_2} and the conventional leakage-based method in \cite{Sadek2007}, respectively, for the first and second desired receivers. From Fig.~\ref{BER_eave_unknown}~(a), similar to Fig.~\ref{BER_eave_known}~(a), we can see that the proposed method still outperforms the OP by approximately an order-of-magnitude in the desired direction $60^\circ$, where the BER value of the proposed method is about one-third of the conventional leakage-based method.  More importantly, the BER of the proposed method is about $40$ percent in the remaining region outside the main-lobe of the desired direction, and the curve fluctuates less than that of the other two methods. This provides an effective barrier to prevent potential eavesdroppers recovering confidential information. Fig.~\ref{BER_eave_unknown}~(b) presents a similar BER performance trend for the second desired direction $120^\circ$ as Fig.~\ref{BER_eave_unknown}~(a).

\begin{figure}[!t]
  \centering
  \includegraphics[width=0.5\textwidth]{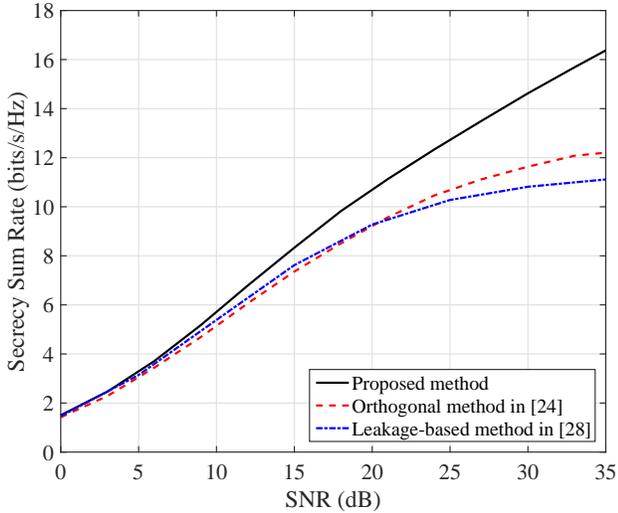}\\
  \caption{Secrecy sum rate versus SNR with imperfect desired and eavesdropper directions.}\label{SSR_eave_known}
\end{figure}

In the following, we will evaluate the performance of the proposed method from the SSR aspect. Fig.~\ref{SSR_eave_known} demonstrates the curves of SSR versus SNR for the proposed method with imperfect desired and eavesdropper directions in Section III-A, the OP method in \cite{Ding2015_2} and the conventional leakage-based method in \cite{Sadek2007}. As can be seen from Fig.~\ref{SSR_eave_known}, with the increase in SNR, the SSRs of all three methods increase continuously and monotonously. The SSR of the proposed method is always larger than that of the remaining two methods. Their SSR values are up to $16.4$ bit/s/Hz, $12.2$ bit/s/Hz and $11.1$ bit/s/Hz at $\text{SNR}=35\text{dB}$, respectively. Hence, the SSR improvement of the proposed method over the OP method and the conventional leakage-based method is significant. Additionally, the SSRs of the three methods approach the same value as SNR goes to $0$ dB. The rate gap between the proposed method and the other two methods grows gradually with the increase in SNR.

\begin{figure}[!t]
  \centering
  \includegraphics[width=0.5\textwidth]{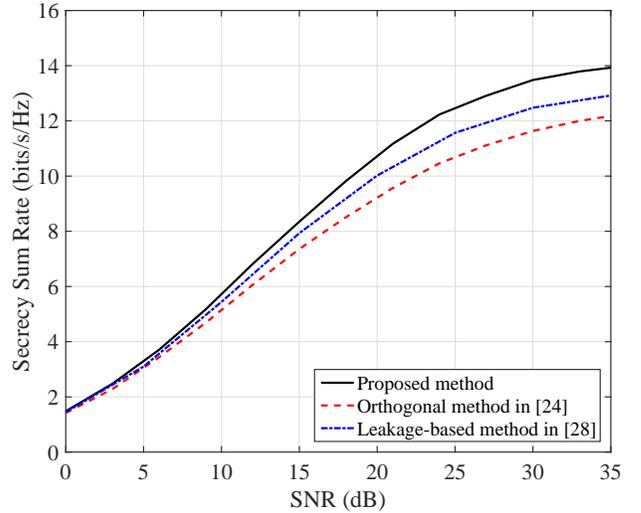}\\
  \caption{Secrecy sum rate versus SNR with unknown eavesdropper directions.}\label{SSR_eave_unknown}
\end{figure}

In the scenario of unknown eavesdropper directions, Fig.~\ref{SSR_eave_unknown} demonstrates the curves of SSR versus SNR for the proposed method in Section III-B, the OP method in \cite{Ding2015_2} and the conventional leakage-based method in \cite{Sadek2007}. As indicated in Fig.~\ref{SSR_eave_unknown}, the proposed method performs better than the OP method and the conventional leakage-based method in terms of SSR, in particular, in the medium and high SNR regions. Their SSR values are up to $14.0$ bit/s/Hz, $12.2$ bit/s/Hz and $12.8$ bit/s/Hz at $\text{SNR}=35~\text{dB}$, respectively,  in which case that the proposed method shows an approximate $15$ and $9$ percent SSR improvement over the OP and the conventional leakage-based method, respectively. Additionally, the SSRs of all three methods tend to be the same in the low SNR region. The rate gap between the proposed method and the other two methods, similar to Fig.~\ref{SSR_eave_known}, grows gradually with the increase in SNR.

Note that the SSRs of the proposed method in Fig.~\ref{SSR_eave_unknown} are less than the corresponding SSRs of the proposed method in Fig.~\ref{SSR_eave_known} in the middle and high regions. This implies that the measured values of eavesdropper direction angles play an important role in improving the SSR performance.

\section{Conclusion}
In this paper, we study robust precoding in MU-MIMO directional modulation systems under imperfect desired directions and eavesdropper directions or unknown eavesdropper directions. To realize the high-performance robust secure multi-message-stream simultaneous transmission in MU-MIMO systems, we propose an MLI-based precoding method. In the first stage, the precoding vector per desired user is optimized to maximize the transmit power of useful confidential messages along the main-lobe of the corresponding desired direction and correspondingly minimize the transmit power of AN along all eavesdropper directions. In the second stage, the AN projection matrix is designed to force AN to all eavesdropper directions with only a small amount of residual AN along the main-lobes of the desired directions. Due to the use of MLI, the proposed method requires no perfect direction knowledge or the distribution of measurement errors of direction angle, and at the same time provides a robust performance. Simulation results verify the performance benefits of our method. Compared with OP, the proposed method can achieve an one-order-magnitude improvement on BER and at the same time its secrecy-sum-rate along the desired directions is shown to have a substantial enhancement over that of OP in the high SNR region. The proposed scheme can be applied to the future satellite communications, mobile communications, D2D, V2V, unmanned-aerial-vehicles networks, and internet of things.

\begin{figure*}[!t]
\normalsize
\setcounter{mytempeqncnt}{\value{equation}}
\begin{align}\label{Simplify_SLNRdk}
\text{MLI-SLNR}(\mathbf{v}_{d_k})
& = \frac{\int_{S_{d_k}}\alpha_1^2\beta_1^2P_s\mathbf{v}_{d_k}^H\mathbf{h}(\theta)\mathbf{h}^H(\theta)\mathbf{v}_{d_k}\,d\theta }{\int_{S_{d_k}}\sigma^2_{d_k}\,d\theta+\sum\limits_{i=1,i \neq k}^K\int_{S_{d_i}}\alpha_1^2\beta_1^2P_s\mathbf{v}_{d_k}^H\mathbf{h}(\theta)\mathbf{h}^H(\theta)\mathbf{v}_{d_k}\,d\theta+\sum\limits_{m=1}^M\int_{S_{e_m}}\alpha_1^2\beta_1^2P_s\mathbf{v}_{d_k}^H\mathbf{h}(\theta)\mathbf{h}^H(\theta)\mathbf{v}_{d_k}\,d\theta}\nonumber\\
& \overset{a}= \frac{\alpha_1^2\beta_1^2P_s\mathbf{v}_{d_k}^H\int_{S_{d_k}}\mathbf{h}(\theta)\mathbf{h}^H(\theta)\,d\theta \mathbf{v}_{d_k} }{\int_{S_{d_k}}\sigma^2_{d_k}\,d\theta+\alpha_1^2\beta_1^2P_s\mathbf{v}_{d_k}^H\sum\limits_{i=1,i \neq k}^K\int_{S_{d_i}}\mathbf{h}(\theta)\mathbf{h}^H(\theta)\,d\theta\mathbf{v}_{d_k}+\alpha_1^2\beta_1^2P_s\mathbf{v}_{d_k}^H\sum\limits_{m=1}^M\int_{S_{e_m}}\mathbf{h}(\theta)\mathbf{h}^H(\theta)\,d\theta\mathbf{v}_{d_k}}\nonumber\\
& \overset{b}= \frac{\alpha_1^2\beta_1^2P_s\mathbf{v}_{d_k}^H\mathbf{R}_{S_{d_k}}\mathbf{v}_{d_k} }{\sigma_{d_k}^2\theta_{BW}+\alpha_1^2\beta_1^2P_s\mathbf{v}_{d_k}^H\sum\limits_{i=1,i \neq k}^K\mathbf{R}_{S_{d_i}}\mathbf{v}_{d_k}+\alpha_1^2\beta_1^2P_s\mathbf{v}_{d_k}^H\sum\limits_{m=1}^M\mathbf{R}_{S_{e_m}}\mathbf{v}_{d_k}}\nonumber\\
& = \frac{\alpha_1^2\beta_1^2P_s\mathbf{v}_{d_k}^H\mathbf{R}_{S_{d_k}}\mathbf{v}_{d_k} }{\sigma_{d_k}^2\theta_{BW}+\alpha_1^2\beta_1^2P_s\mathbf{v}_{d_k}^H\mathbf{R}_{S_d\backslash S_{d_k}}\mathbf{v}_{d_k}+\alpha_1^2\beta_1^2P_s\mathbf{v}_{d_k}^H\mathbf{R}_{S_e}\mathbf{v}_{d_k}}\nonumber\\
& = \frac{\mathbf{v}_{d_k}^H\mathbf{R}_{S_{d_k}}\mathbf{v}_{d_k} }{\frac{\sigma_{d_k}^2\theta_{BW}}{\alpha_1^2\beta_1^2P_s}+\mathbf{v}_{d_k}^H\mathbf{R}_{S_d\backslash S_{d_k}}\mathbf{v}_{d_k}+\mathbf{v}_{d_k}^H\mathbf{R}_{S_e}\mathbf{v}_{d_k}}\nonumber\\
& = \frac{\mathbf{v}_{d_k}^H\mathbf{R}_{S_{d_k}}\mathbf{v}_{d_k}}{\mathbf{v}_{d_k}^H\left(\frac{\sigma_{d_k}^2\theta_{BW}}{\alpha_1^2\beta_1^2P_s}\mathbf{I}_N+\mathbf{R}_{S_d\backslash S_{d_k}}+\mathbf{R}_{S_e}\right)\mathbf{v}_{d_k}}.
\end{align}
\setcounter{equation}{\value{mytempeqncnt}}
\hrulefill
\vspace*{4pt}
\end{figure*}

\appendices
\section{Simplification of $\text{MLI-SLNR}(\mathbf{v}_{d_k})$}
\emph{Proof:} The simplification of $\text{MLI-SLNR}(\mathbf{v}_{d_k})$ is shown in (\ref{Simplify_SLNRdk}), where $\overset{a}{=}$ is achieved by extracting integral term $\mathbf{h}(\theta)\mathbf{h}^H(\theta)$, $\overset{b}{=}$ is achieved by replacing $\int_{S_{d_k}}\mathbf{h}(\theta)\mathbf{h}^H(\theta)\,d\theta$, $\int_{S_{d_i}}\mathbf{h}(\theta)\mathbf{h}^H(\theta)\,d\theta$ and $\int_{S_{e_m}}\mathbf{h}(\theta)\mathbf{h}^H(\theta)\,d\theta$ by $\mathbf{R}_{S_{d_k}}$, $\mathbf{R}_{S_{d_i}}$ and $\mathbf{R}_{S_{e_m}}$, respectively. \hfill$\blacksquare$

\section{Derivation of $\mathbf{R}_S$}
\emph{Proof:} The integral matrix $\mathbf{R}_{S}$ in (\ref{R_S}) is an $N \times N$ matrix. Here we assume that set $S$ is a union of $I$ separate subintervals as follows
\begin{equation}\label{S_I}
S=\bigcup\limits_{i=1}^I S_i,
\setcounter{equation}{42}
\end{equation}
where
\begin{equation}\label{S_i}
S_i=[\theta_\text{min}^i,~ \theta_\text{max}^i].
\end{equation}
Then the $(p,q)$ entry of $\mathbf{R}_{S_i}$  has the following form
\begin{align}\label{R_Si_pq}
\mathbf{R}_{S_i}(p,q)
& = \int_{\theta_\text{min}^i}^{\theta_\text{max}^i}\mathbf{h}_p(\theta)\mathbf{h}_q^H(\theta)\,d\theta\nonumber\\
& = \int_{\theta_\text{min}^i}^{\theta_\text{max}^i}\frac{1}{\sqrt{N}}e^{\frac{j2\pi(p-(N+1)/2)d\cos\theta}{\lambda}}\nonumber\\ &\cdot\frac{1}{\sqrt{N}}e^{-\frac{j2\pi(q-(N+1)/2)d\cos\theta}{\lambda}}\,d\theta\nonumber\\
& = \frac{1}{N}\int_{\theta_\text{min}^i}^{\theta_\text{max}^i}e^{\frac{j2\pi(p-q)d\cos\theta}{\lambda}}\,d\theta.
\end{align}
Let us define the center point and length of integral interval as
\begin{align}\label{theta0_i}
\theta_0^i = & \frac{\theta_{\text{min}}^i+\theta_{\text{max}}^i}{2},
\end{align}
and
\begin{align}\label{Delta-theta_i}
\Delta\theta_i = & \theta_{\text{max}}^i-\theta_{\text{min}}^i,
\end{align}
respectively, and the new integral variable is
\begin{equation}\label{x_sub}
x=\frac{2\pi}{\Delta\theta_i}(\theta-\theta_0^i),
\end{equation}
i.e.,
\begin{equation}\label{theta_sub}
\theta=\frac{\Delta\theta_i}{2\pi}x+\theta_0^i.
\end{equation}
Using the above definition or transformation, Eq. (\ref{R_Si_pq}) is rewritten as
\begin{align}\label{R_S_pq2}
& \mathbf{R}_{S_i}(p,q)\nonumber\\
& = \frac{1}{N}\int_{-\pi}^{\pi}e^{\frac{j2\pi(p-q)d\cos(\frac{\Delta\theta_i}{2\pi}x+\theta_0^i)}{\lambda}}\cdot\frac{\Delta\theta_i}{2\pi}\,dx\nonumber\\
& =
\frac{\Delta\theta_i}{2\pi N}\int_{-\pi}^{\pi}e^{\frac{j2\pi(p-q)d\cos(\frac{\Delta\theta_i}{2\pi}x+\theta_0^i)}{\lambda}}\,dx\nonumber\\
& =
\frac{\Delta\theta_i}{2\pi N}\int_{-\pi}^{\pi}e^{\frac{j2\pi(p-q)d\left[\cos(\theta_0^i)\cos(\frac{\Delta\theta_i}{2\pi}x)-\sin(\theta_0^i)\sin(\frac{\Delta\theta_i}{2\pi}x)\right]}{\lambda}}\,dx\nonumber\\
& =
\frac{\Delta\theta_i}{2\pi N}\int_{-\pi}^{\pi}e^{\left[a_{pq}^i\cos(c_ix)+b_{pq}^i\sin(c_ix)\right]}\,dx\nonumber\\
& = \frac{\Delta\theta_i}{N}\textrm{g}_B(a_{pq}^i,b_{pq}^i,c_i),
\end{align}
where $\textrm{g}_B(\cdot)$ is the extension of the modified Bessel function of the first kind with the integer order $0$ \cite{Gradshteyn2007}, i.e.,
\begin{equation}\label{gB}
\textrm{g}_B(a_{pq}^i,b_{pq}^i,c_i)=\frac{1}{2\pi}\int_{-\pi}^{\pi}e^{\left[a_{pq}^i\cos(c_ix)+b_{pq}^i\sin(c_ix)\right]}\,dx,
\end{equation}
\begin{equation}\label{ai_expre}
a_{pq}^i \triangleq \frac{j2\pi(p-q)d\cos(\theta_0^i)}{\lambda},
\end{equation}
\begin{equation}\label{bi_expre}
b^i_{pq} \triangleq -\frac{j2\pi(p-q)d\sin(\theta_0^i)}{\lambda},
\end{equation}
and
\begin{equation}\label{ci_expre}
c_i \triangleq \frac{\Delta\theta_i}{2\pi}.
\end{equation}
This completes the derivation of the $(p,q)$ entry $\mathbf{R}_{S_i}(p,q)$ of matrix $\mathbf{R}_{S_i}$. Note that set $S$ is a union of $I$ separated subintervals as shown in (\ref{S_I}). Then the $(p,q)$ element of matrix $\mathbf{R}_S$ is expanded as the summation of integrations over all subintervals
\begin{equation}\label{R_S_K}
\mathbf{R}_S(p,q)=\sum\limits_{i=1}^I\mathbf{R}_{S_i}(p,q)=\sum\limits_{i=1}^I\frac{\Delta\theta^i}{N}\textrm{g}_B(a_{pq}^i,b_{pq}^i,c_i).
\end{equation}
This completes the derivation of matrix $\mathbf{R}_S$.\hfill$\blacksquare$



\ifCLASSOPTIONcaptionsoff
  \newpage
\fi

\bibliographystyle{IEEEtran}

\bibliography{IEEEfull,Integration_MU_DM}

\end{document}